**Covalently bound Bovine Serum Albumin (BSA) protein modified Hydrogenated Diamond Like Carbon (HDLC) surface as biosensor application**

**Hari Shankar Biswas[a,b],** Kausik Gupta[c], Jagannath Datta[d], Uday Chand Ghosh[c], Nihar Ranjan Ray[a*]

[a]Nanocrystalline Diamond like Carbon Synthesis Laboratory, Saha Institute of Nuclear Physics 1/AF, Bidhan Nagar, Kolkata-700 064 India

[b] Department of Chemistry, Surendranath College, 24/2 Mahatma Gandhi road, Kolkata-700009, India

[c]Department of Chemistry, Presidency University, 86/1 College Street, Kolkata-700 073 , India

[d]Analytical Chemistry Division, Bhabha Atomic Research Centre, Variable Energy Cyclotron Centre, 1/AF, Bidhan Nagar, Kolkata-700 064, India

*Corresponding author

**Abstract**

We have developed a biosensor based on BSA with the help of metal ions binding mechanism to detect and remove inorganic As (III), Cu (II), Pb (II) from water like fishing by hooking system. *BSA was immobilized via DOPA onto the HDLC (named as BD-HDLC) surface having high loading capacity of proteins with no conformal change and having covalent interaction.* These BD-HDLC samples, separately was immersed in the 25ml $1 \times 10^{-5}$ M metal ions solution in Tris buffers at pH 6.5, for 2 h at room temperature, under continuous slow stirring conditions using a magnetic stirrer. The metal ions treated BD-HDLC film was rinsed by ultrapure water and dried with nitrogen. We have studied that BD-HDLC surface shows high binding ligand activity to the potential binding site of $Cu^{2+}$, $Pb^{2+}$, and $As^{3+}$. Additionally, BD-HDLC−$Cu^{2+}$, BD-HDLC-$Pb^{2+}$ and BD-HDLC−$As^{3+}$ complexes can be used to study their individual interactions with each other. Thus, we demonstrate that excellent BD-HDLC can be used as a metal ion sensor at pH 6.5 upto $1 \times 10^{-5}$ M concentration of metal ion solution.

## 1. Introduction

It is well known that Lead is an important trace element in humans; in cattle, Copper deficiency is a common issue and Arsenic is an environmental pollutant and high levels of Arsenic can cause a wide range of health effects[1], including cancers of the bladder, lung, skin, and kidney. Also the interactions of Copper, Arsenic and Sulfur in ruminants have the adverse effect not only on the rumen of grazing animals and but also for all living organisms due to their inhibitory action on enzymes [2]. Thus, the detection of Lead, Arsenic, and Copper etc. in foodstuffs, medications, and the environment at low levels are necessary for health protection but the mechanism of these hazardous effects of Lead, Arsenic, and Copper etc. remains in dark. The only possible mechanism [3] of toxicity is the metal ions binding to cellular proteins and the identification of such binding analytically challenging. BSA protein which contains a single free sulfhydryl group of residue Cys34 per molecule and which thus represents a macromolecular thiol[1] compound and can bind to trivalent [4] Arsenic, bi-valent Lead[2]. Since currently, biosensors [5-6] have been gained interest to detect metal ions, we have developed a biosensor based on BSA with the help of metal ions binding mechanism to detect and remove inorganic As (III), Cu (II), Pb (II) from water like fishing by hooking system.

Oblak et al. (2004) studied the serum albumin contains 585 amino acids, (Table 1) of which 35 are cysteine and cystines where Only the first cysteine in the sequence remains a free thiol while the rest participate in the formation of 17 disulfide bonds and Peters (1996) observed that these characteristics of the cysteine residues are conserved in the serum albumin of all vertebrates. We have demonstrated earlier [7] protein immobilization method onto HDLC surface *having high loading capacity of proteins with no conformal change and having covalent interaction*. Now if conformation of BSA protein in BD-HDLC remains intact then it can act as ligand to metal ions in solution.

A detailed knowledge of BD-HDLC structure of this protein is imperative to understand its physical properties, and binding modes with $Cu^{2+}$, $Pb^{2+}$ and $As^{3+}$. Once the BD-HDLC structure is

Table (1) Aminoacid composition of Serum Albumin of all vertebrate

| Amino acid composition of BSA (Brown, 1975; Patterson and Geller, 1977; McGillivray *et al.*, 1979; Hirayama et al., 1990); Reed *et al.*, 1980. | | | |
|---|---|---|---|
| Ala48 | Cys 35 | Asp 41 | Glu 58 |
| Phe 30 | Gly 17 | His 16 | Ile 15 |
| Lys60 | Leu 65 | Met 5 | Asn 14 |
| Pro 28 | Gln 21 | Arg 26 | Ser 32 |
| Thr 34 | Val 38 | Trp 3 | Tyr 21 |

known, it can act as a template to investigate the interactions that occur among $Cu^{2+}$, $Pb^{2+}$, $As^{3+}$ and BD-HDLC. Excellent BD-HDLC surface that produce high binding ligand activity are essential to achieve this study to simulate the potential binding site of $Cu^{2+}$, $Pb^{2+}$, and $As^{3+}$, BSA and their binding characteristic in the bovine ruminant system. Additionally, BD-HDLC−$Cu^{2+}$,

BD-HDLC-$Pb^{2+}$ and BD-HDLC−$As^{3+}$ complexes can be used to study their individual interactions with each other.

## 2. Experimental

BD-HDLC samples, separately was immersed in the 25ml $1\times10^{-5}$ M $Pb(NO_3)_2$ solution in Tris buffers at pH 6.5, 25ml $1\times10^{-5}$ M $Cu(NO_3)_2$ solution in Tris buffers at pH 6.5 and 25ml $1\times10^{-5}$ M $NaAsO_2$ solution in Tris buffers at pH 6.5, for 2 h at room temperature, under continuous slow stirring conditions using a magnetic stirrer. The metal ions treated BD-HDLC film was rinsed by ultrapure water and dried with nitrogen.

## 3. Characterization

### 3.1 Characterization by FTIR using ATR Accessories

Fourier transform infrared spectroscopy (FTIR) technique is one of the very good characterization techniques for protein conformational analysis that can be applied for structural of immobilized proteins onto solid HDLC surface. Here we are using this technique to obtain information on protein structure on BD-HDLC surface before and after interaction with some metal ions [8]. In best of my knowledge there is no report on Second derivative [9-10] FTIR spectra for investigation conformational change of immobilized protein onto solid substrate. For BSA protein the amide I frequency range has been assigned as α-helix 1655$cm^{-1}$, β-sheet and turn 1678 and 1633$cm^{-1}$ [11]. Fig.1 is the FTIR spectra of HDLC [7], D-HDLC [7], BD-HDLC [7], BD-HDLC-$Cu^{2+}$, BD-HDLC–$Pb^{2+}$, BD-HDLC-$As^{3+}$ and Fig.2 (a,b,c and d) indicates the second derivative FTIR of BD-HDLC-$Cu^{2+}$, BD-HDLC–$Pb^{2+}$, BD-HDLC-$As^{3+}$. In BD-HDLC in Fig.2a the amide I region 1600$cm^{-1}$, 1612$cm^{-1}$, 1629$cm^{-1}$, 1675$cm^{-1}$ are present but absence of α-helix 1655$cm^{-1}$. This indicates it is engaged for covalent binding with (DOPA treated HDLC) D-HDLC and band 1675 and 1628cm$^{-1}$ can be assigned to β-sheet and turn segment of BSA [12]. In presence of Cu$^{2+}$ ion, the intensity of the amide I band at 1671cm$^{-1}$, 1637cm$^{-1}$, 1624cm$^{-1}$ changes due to increase β-sheet and turn structures. In presence of Pb$^{2+}$ ion intensity at 1634cm$^{-1}$ and 1680 cm$^{-1}$ are strong and indicates the change of β-sheet and turn structures upon Pb$^{2+}$ ion binding .This result confirms that the covalently attached BSA molecule undergoes conformational changes from α-helix to β-sheet [12] . So Pb$^{2+}$ and Cu$^{2+}$ ion bind through protein carbonyl groups as well as through C-N group can be said from the positive feature of 1555cm$^{-1}$. Quagraine et al. explained the N-terminal region of bovine serum albumin (Asp-Thr-His-Lys) is known to provide a specific binding site[13] for Cu$^{2+}$ ions, with the histidine residue thought to be mainly responsible for specificity also Yu et al. described the copper binding ability of bovine serum albumin and human serum albumin are more or less same[14]. But for As$^{3+}$ion do not increase the β-sheet it is mainly bind with amide II of C-N group and thionyl group.

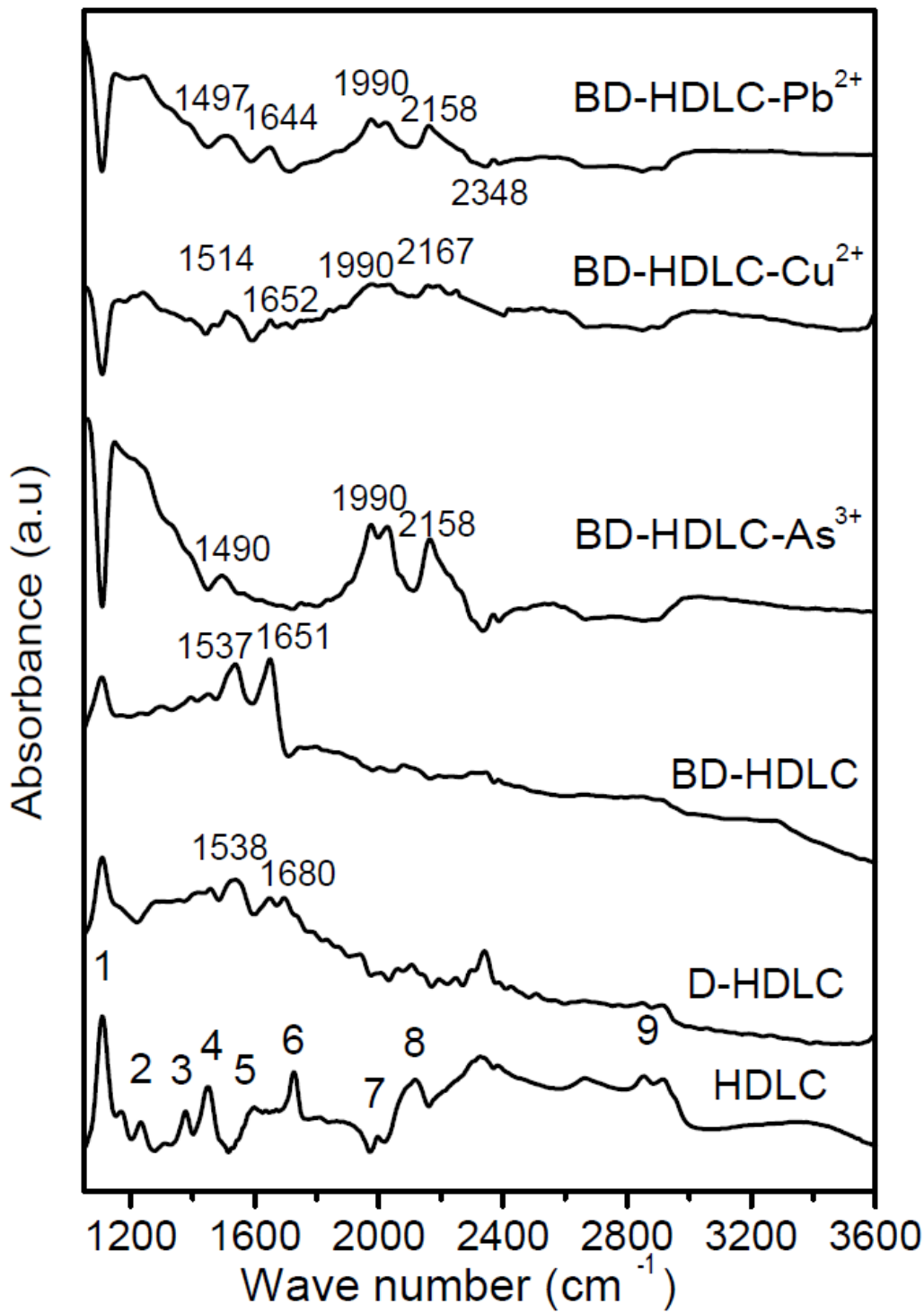

**Fig.1** FTIR spectra HDLC and modified HDLC by diamond ATR

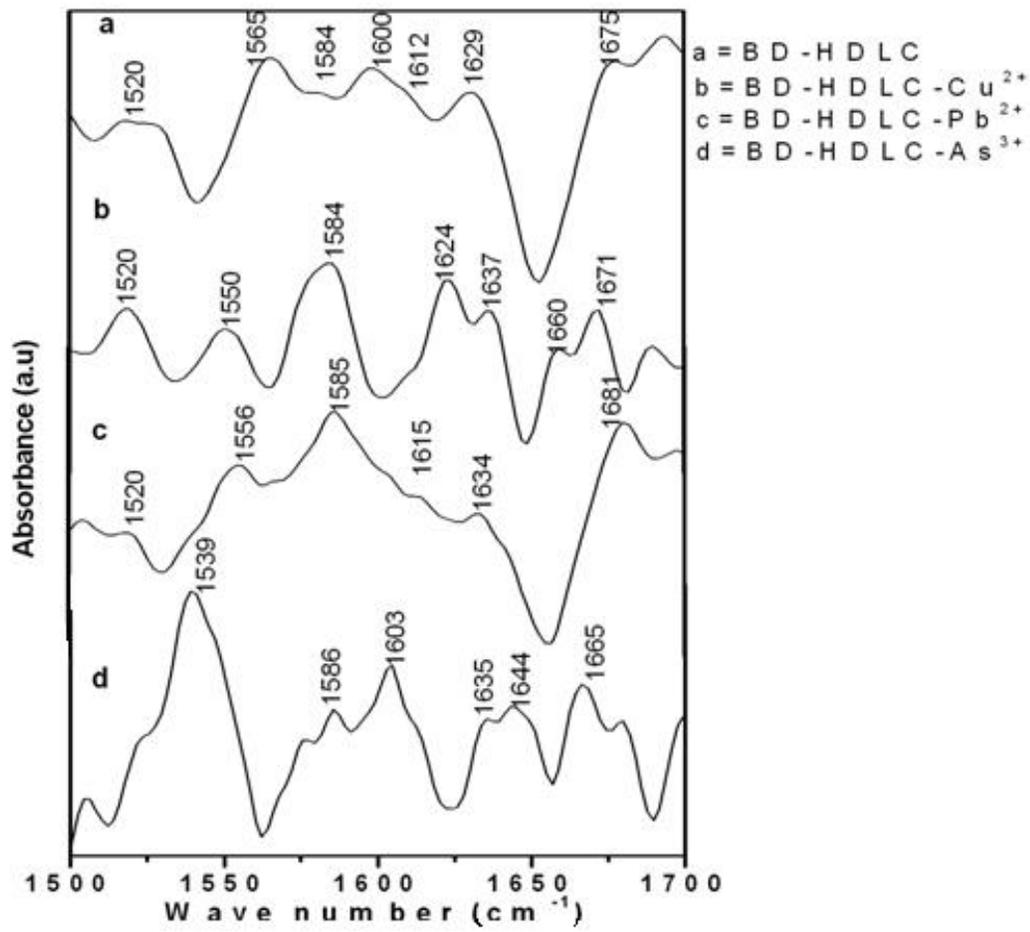

**Fig.2** Second derivative of FTIR spectrum a) BD-HDLC, b) BD-HDLC-$Cu^{2+}$, c) BD-HDLC-$Pb^{2+}$, d) BD-HDLC-$As^{3+}$.

## 3.2 Study by AFM

It is a powerful characterization technique which capable of revealing surface structures with superior spatial resolution to asses particle size and local surface topography. Fig.38 represent 3D images of BD-HDLC (image a), BD-HDLC-$Pb^{2+}$ (image b), BD-HDLC-$As^{3+}$ (image c) and BD-HDLC-$Cu^{2+}$. Image-a showed continuous covalently bound BSA protein surface morphology. Image-b displayed cage-like complex structure manifested by the change of protein conformation to composite formation which is supported by FTIR study. Image -c showed $As^{3+}$-BSA complex in 3D view with metal chelate complex like structure. Ngu et.al (1996) showed that $As^{3+}$ bound to three sulfur atoms of Human Metallothionein to form trigonal pyramidal coordination complex where the sulfur atoms act as terminal ligand. They also reported trialkyl trithioarsenite [$As(SR)_3$] compound having trigonal-pyramidal geometry[15]. Image-d displayed the cage-like structure of BD-HDLC-$Cu^{2+}$ coordinate complex also supported by FTIR study. These different metal ion-BD-HDLC complexes show different surface morphology with various binding sites.

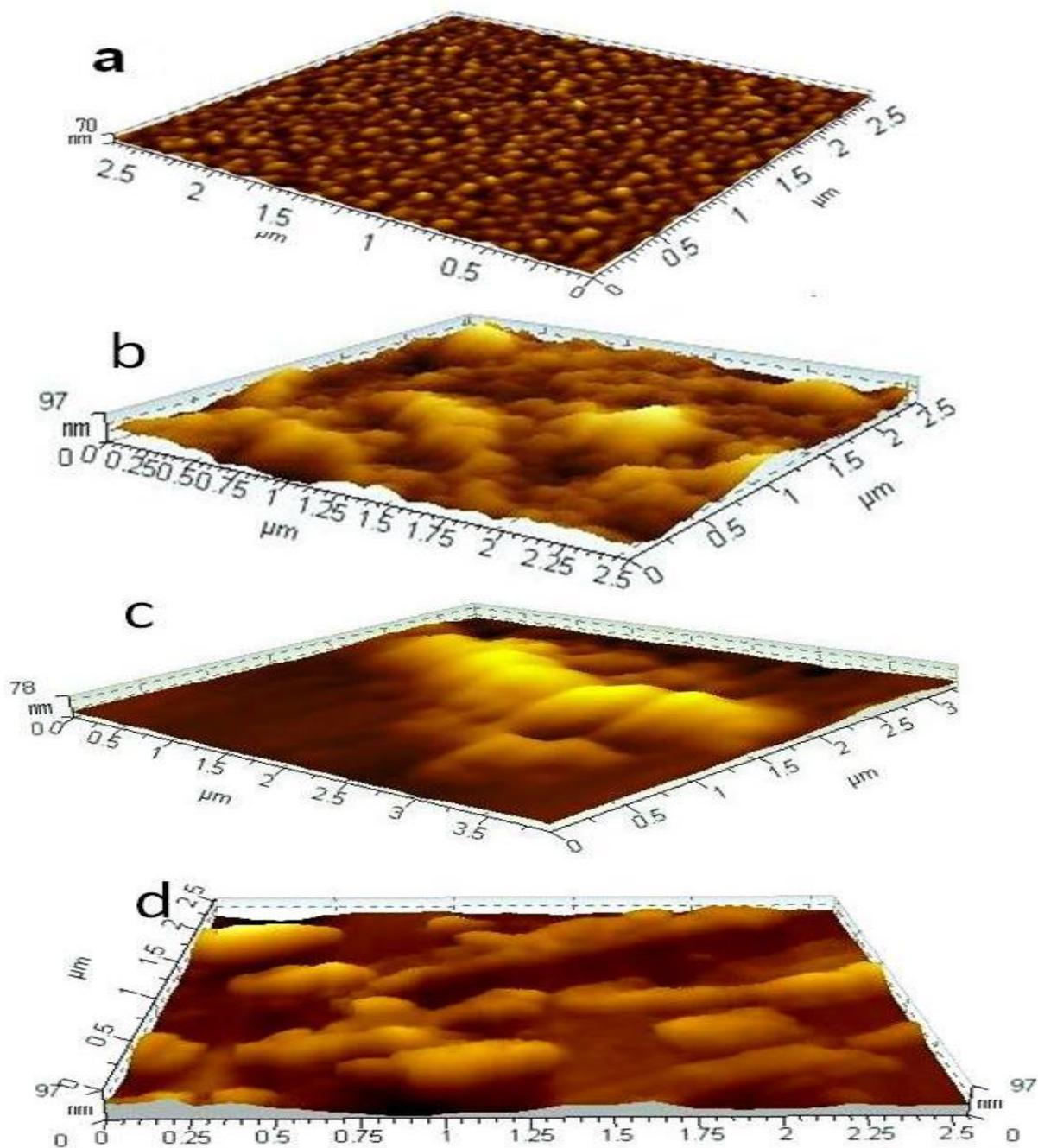

**Fig.3** Non-Contact mode AFM topography images of (a) BSA treated HDLC (BD-HDLC) surface: scan size 1.5 μm × 1.5 μm (b) $Pb^{2+}$ ions solution treated BD-HDLC surface: scan size 2.5 μm × 2.5 μm. (c) $As^{3+}$ ions solution treated BD-HDLC surface: scan size 2.5 μm × 2.5 μm. (d) $Cu^{2+}$ ions solution treated BD-HDLC surface: scan size 2.5 μm × 2.5 μm.

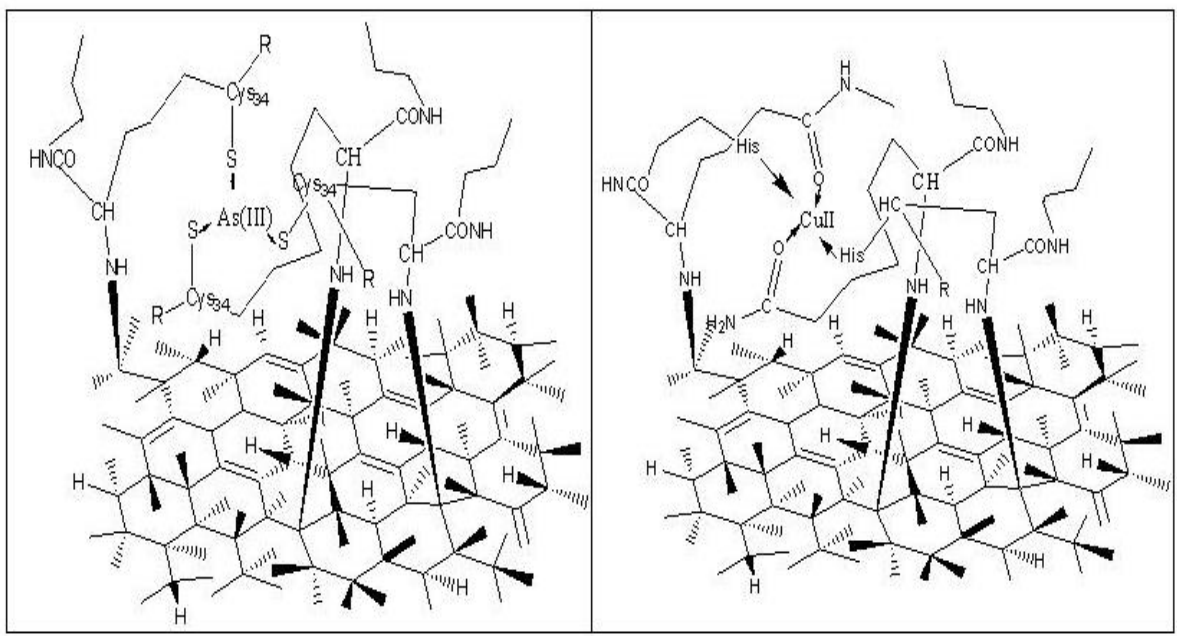

**Fig.4** Schematic model of cage-like structure of a) As-BD-HDLC, b) Cu-BD-HDLC

**Conclusions**

Our studies conducted in this article shows that the covalently bound BSA protein onto HDLC is very suitable as solid state **biosensor** for some heavy metal ions determination in liquid metal ions solution. There are several advantages **to use this as a sensor** since it is a solid state material and easily can made, low cost, sensibility and easy to handle.